# Bound-state solutions of the Schrödinger equation for two novel potentials

A. D. Alhaidari[(a)] and I. A. Assi[(b)]

[(a)] *Saudi Center for Theoretical Physics, P.O. Box 32741, Jeddah 21438, Saudi Arabia*

[(b)] *Department of Physics and Physical Oceanography, Memorial University of Newfoundland, St. John's, Newfoundland & Labrador, A1B 3X7, Canada*

**Abstract:** We solve the one-dimensional Schrödinger equation for the bound states of two potential models with a rich structure as shown by their "spectral phase diagram". These potentials do not belong to the well-known class of exactly solvable problems. The solutions are finite series of square integrable functions written in terms of the Jacobi polynomials.

## 1. Introduction

Fundamental interactions in nature are very few. In fact, we know only of four that may even be unified (merged) into fewer interactions as the energy scale becomes very large. In quantum mechanics these fundamental interactions are modeled in simple systems by even fewer number of potential functions (e.g., $r^{-1}$ for the Coulomb and Kepler problems, $r^2$ for quarks interaction, etc.). However, for complex systems (e.g., those with a large number of constituents), the fundamental interactions become intractable and modelling by simple potential functions becomes non-trivial to almost impossible. However, potential models formed using various functions that satisfy basic physical constraints can give a very good description of certain aspects of the system. For example, the binding of some molecules can be described to an extremely high accuracy by the Morse potential with a proper choice of parameters. Consequently, the search for potential functions that can model the structure and dynamics of various physical systems started since the early conception of quantum mechanics and still continues. Out of all of these models, the interesting ones are those that can be solved exactly for the whole energy spectrum or for a finite portion thereof. The latter solution is referred to as "quasi-exact". Nonetheless, the number of such potential functions is very small. Researchers continue to develop methods to enlarge the class of exactly solvable potential models.

In this work, we use the tridiagonal representation approach (TRA) [1] to obtain exact bound-state solutions of the Schrödinger equation for the following four-parameter potential models

$$V_1(x) = \frac{2}{x^2 + 2a^2}\left[\frac{a^2 A}{x^2} - \frac{a^2 B}{x^2 + 2a^2} + C\right] = \frac{A}{x^2} + \frac{2C - A}{x^2 + 2a^2} - \frac{2a^2 B}{\left(x^2 + 2a^2\right)^2},$$ (1a)

$$V_{\text{II}}(x) = \frac{1}{x(x+2a)}\left[\frac{a^2 A}{x(x+2a)} - \frac{a^2 B}{(x+a)^2} + C\right] = \frac{A/4}{x^2} + \frac{2C - 2B - A}{2x(x+2a)} + \frac{A/4}{(x+2a)^2} + \frac{B}{(x+a)^2}, \quad (1b)$$

where $x \geq 0$. The parameters $\{A, B, C\}$ are dimensionless, real and positive. The scale parameter $a$ is positive with a dimension of length. These two potentials do not belong the known class of exactly solvable problems. Nonetheless, potential $V_{\text{I}}(x)$ was treated in section III.A.6 of Ref. [2] and in section 2.4 of Ref. [3]. However, no exact TRA solutions were obtained because the matrix representation of the wave operator in the chosen basis therein was not tridiagonal despite that the Hamiltonian matrix is. Potential $V_{\text{I}}(x)$ is inverse-squared singular at the origin with a singularity strength $A$. On the other hand, $V_{\text{II}}(x)$ has an inverse-linear as well as an inverse-squared singularity at the origin with respective strength of $(2C - 2B - A)/4a$ and $A/4$. Moreover, both potentials vanish at infinity. The upper bound on the number of bound states is obtained by evaluating the integral $\int_0^\infty xV^-(x)dx$, where $V^-(x) = -V(x)\theta[-V(x)]$ and $\theta(x)$ is the usual step function [4,5]. For $C > 0$, the limits of integration 0 and $\infty$ become $x_\pm$, which are finite, and the value of the integral is also finite. However, for $C < 0$, the upper limit $x_+$ is infinite and the integral diverges. Consequently, with $C > 0$ the two potentials could support a mix of finite number of bound states and resonances. Whereas, if $C < 0$ then they could support an infinite number of bound states. The proper values of the potential parameters for supporting such structures must produce one or two positive real roots for the cubic equation that results from the condition $(dV/dx)_{x_0} = 0$ with $x_0 > 0$. A necessary (but may not be sufficient) condition for the existence of bound states is that $V(x_0) < 0$. The existence of two different real positive roots of the cubic equation implies the possibility of resonances. The cubic equation associated with $V_{\text{I}}(x)$ and $V_{\text{II}}(x)$ are

$$Cs^3 + 2(C - B + A)s^2 + 6As + 4A = 0, \quad (2a)$$

$$Ct^3 + 2(C - B + A)t^2 + (C - B + 4A)t + 2A = 0, \quad (2b)$$

respectively, where $s = (x/a)^2$ and $t = [(x/a) + 1]^2 - 1$. Since $A$ and $C$ are positive, then Descartes' rule of signs [6] for Eq. (2a) and Eq. (2b) dictates that $B$ must be greater than $A + C$ resulting in two positive real roots. Thus, the spectrum will then consist of a mixture of finite number of bound states and resonances. On the other hand, if we had taken $C$ negative, then Descartes' rule of signs would have implied a single positive real root for Eq. (2a) and Eq. (2b) resulting in an infinite number of pure bound states (without resonances). Figure 1 shows several plots of the two potential functions (in units of $A/a^2$) for a fixed value of the parameter ratio $C/A$ and for different values of $B/A$. Figure 2 is the spectral phase diagram (SPD) for the two potentials showing the distribution of their corresponding energy spectrum (scattering states, bound states, and resonances) as a function of the potential parameters. Note that the observations made above about the sign of the potential parameter $C$ is consistent with the SPDs shown in the figure. A detailed description of the SPD, its benefits and how to construct it are found in Ref. [7].

In the TRA [1,2], the solution of the Schrödinger equation, $\mathcal{D}\psi(x) = 0$, is written as a bounded convergent series of discrete square-integrable functions $\{\phi_n\}$. That is, $\psi(x) = \sum_n f_n \phi_n(y)$, where $y = y(x)$ is a coordinate transformation and $\{f_n\}$ are the expansion coefficients. The basis set $\{\phi_n\}$ must be complete and should result in a tridiagonal matrix representation for the wave operator, $\langle \phi_n | \mathcal{D} | \phi_m \rangle$. That is, the action of the wave operator on the basis element should read [1]

$$\mathcal{D}\phi_n(y) = W(y)\left[d_n \phi_n(y) + b_{n-1}\phi_{n-1}(y) + c_n \phi_{n+1}(y)\right], \tag{3}$$

where $W(y)$ is a node-less entire function and $\{b_n, c_n, d_n\}$ are constant coefficients. Moreover, the integral $\int_{x_-}^{x_+} \phi_m(y) W(y) \phi_n(y) dx$ must be proportional to $\delta_{m,n}$. Hence, the wave equation $\mathcal{D}\psi(x) = 0$ becomes a three-term recursion relation for the expansion coefficients $\{f_n\}$ as follows

$$d_n F_n + b_{n-1} F_{n-1} + c_n F_{n+1} = 0, \tag{4}$$

where we have written $f_n = f_0 F_n$ making $F_0 = 1$. Accordingly, the solution of the wave equation, $\mathcal{D}\psi(x) = 0$, reduces to an algebraic solution of the discrete equation (4). Moreover, the set $\{f_n\}$ contains all physical information about the system modelled by the potential.

Now, to solve for the continuous spectrum or for an infinite discrete spectrum of a given physical system, completeness of the set $\{\phi_n\}$ implies that it is an infinite and dense set. However, for systems with a finite number of bound states, a finite basis set $\{\phi_n\}_{n=0}^{N}$ could produce a faithful representation of the physical system provided that the number of bound states is less than or equal to the size of the basis $N+1$. Additionally, for quasi-exact solution where one looks for a finite portion of the infinite discrete spectrum, such finite basis set could lead to a very good approximation for that portion of the spectrum with an accuracy that increases as $N$ does.

For the finite bound states of the system modelled by either $V_I(x)$ or $V_{II}(x)$, we choose a finite basis set with the following elements

$$\phi_n(y) = (y-1)^\alpha (y+1)^{-\beta} Q_n^{(\mu,\nu)}(y), \tag{5}$$

where $Q_n^{(\mu,\nu)}(y)$ is the Jacobi polynomial defined on the semi-infinite real line $y(x) \geq 1$ as shown in Appendix A. The real basis parameters $\{\alpha, \beta, \mu, \nu\}$ are to be determined below in terms of the physical parameters $\{a, A, B, C\}$ by the TRA constraints. Moreover, $n = 0, 1, .., N$ with $N = \left\lfloor -\frac{\mu+\nu+1}{2} \right\rfloor$ where $\lfloor z \rfloor$ stands for the largest integer less than $z$.

In the atomic units $\hbar = m = 1$, the time-independent Schrödinger equation, $\mathcal{D}\psi(x) = 0$, in the configuration space $x$ for the potential $V(x)$ and energy $E$ reads as follows:

$$\left[ -\frac{1}{2}\frac{d^2}{dx^2} + V(x) - E \right]\psi(x) = 0. \tag{6}$$

In sections 2 and 3, we use the TRA to solve this equation for the two potential models, respectively. We conclude in section 4 with some remarks and discussion of our findings.

## 2. TRA solution of the potential model (1a)

In this case, we choose the coordinate transformation $y(x) = (x/a)^2 + 1$. Writing the differential operator $\frac{d^2}{dx^2}$ and the potential $V_1(x)$ in terms of the dimensionless variable $y$, Eq. (6) becomes

$$\mathcal{D}\psi(x) = -\frac{2/a^2}{y+1}\left[(y^2-1)\frac{d^2}{dy^2} + \frac{y+1}{2}\frac{d}{dy} - \frac{A}{y-1} + \frac{B}{y+1} - C + \frac{\varepsilon}{2}(y+1)\right]\psi(x) = 0, \quad (7)$$

where $\varepsilon = a^2 E$. To solve this equation, we substitute the series $\psi(x) = \sum_n f_n \phi_n(y)$. Consequently, we need to evaluate the action of the wave operator on the basis elements, $\mathcal{D}|\phi_n\rangle$, and then impose the TRA constraint (3). To achieve that we choose the basis parameters $2\alpha = \mu + \frac{1}{2}$ and $2\beta = -\nu - 1$ then use the differential equation of the Jacobi polynomial $Q_n^{(\mu,\nu)}(y)$ given in Appendix A by (A2) to obtain

$$\mathcal{D}|\phi_n(y)\rangle = -\frac{1}{a^2}\frac{(y-1)^\alpha}{(y+1)^{\beta+1}}\left\{\frac{\mu^2 - \frac{1}{4}}{y-1} - \frac{\nu^2 - 1}{y+1} + \frac{1}{2}\left[(2n+\mu+\nu+1)^2 - \frac{1}{4}\right]\right.$$
$$\left. -\frac{2A}{y-1} + \frac{2B}{y+1} - 2C + \varepsilon(y+1)\right\}Q_n^{(\mu,\nu)}(y) \quad (8)$$

Now, the TRA constraint (3) and the recursion relation of the Jacobi polynomials (A3) dictate that terms inside the curly brackets in (8) must be linear in $y$. Thus, we should choose the Jacobi polynomial parameters as

$$\mu^2 = 2A + \tfrac{1}{4}, \qquad \nu^2 = 2B + 1. \quad (9)$$

Reality dictates that $A \geq -\frac{1}{8}$ and $B \geq -\frac{1}{2}$. These constraints are automatically satisfied since we already required that $A > 0$ and $B > 0$. Moreover, the polynomial parameters inequalities $\mu > -1$ and $\mu + \nu < -2N - 1$ dictate that $\mu = \sqrt{2A + \frac{1}{4}}$ and $\nu = -\sqrt{2B+1}$. It also shows that the maximum number of bound states that could be obtained by our TRA solution, which is $N+1$, becomes $\left\lfloor \frac{1}{2}\left(\sqrt{2B+1} - \sqrt{2A+\frac{1}{4}} - 1\right)\right\rfloor + 1$. With these choices of basis parameters, Eq. (8) becomes

$$\mathcal{D}|\phi_n\rangle = -\frac{1}{a^2}\frac{(y-1)^\alpha}{(y+1)^{\beta+1}}\left\{\frac{1}{2}\left[(2n+\mu+\nu+1)^2 - \frac{1}{4}\right] + \varepsilon - 2C + \varepsilon y\right\}Q_n^{(\mu,\nu)}(y). \quad (10)$$

Using the three-term recursion relation for the Jacobi polynomials (A3) in this equation and comparing the result to the TRA constraint (3), we obtain

$$W(y) = \frac{-E}{y+1}, \quad (11a)$$

$$d_n = \left\{\frac{1}{2}\left[(2n+\mu+\nu+1)^2 - \frac{1}{4}\right] + \varepsilon - 2C\right\}\frac{1}{\varepsilon} + \frac{\nu^2 - \mu^2}{(2n+\mu+\nu)(2n+\mu+\nu+2)}, \quad (11b)$$

$$b_n = \frac{2(n+1)(n+\mu+\nu+1)}{(2n+\mu+\nu+1)(2n+\mu+\nu+2)}, \quad c_n = \frac{2(n+\mu+1)(n+\nu+1)}{(2n+\mu+\nu+2)(2n+\mu+\nu+3)}. \quad (11c)$$

For $n = 0,1,..,N$ with $N = \left\lfloor -\frac{\mu+\nu+1}{2} \right\rfloor$, we can show that $b_n c_n > 0$ for all $n \leq N$. Then according to Favard theorem [8] (a.k.a. the spectral theorem; see section 2.5 in [9]) the sequence $\{F_n(z)\}$ satisfying the three-term recursion relation (4) forms a set of orthogonal polynomials with $f_0^2(z)$ being the positive definite weight function. The polynomial argument $z$ depends on the energy $\varepsilon$ and the potential parameter $C$. If we define the polynomial

$$P_n = \frac{(\mu+1)_n (\nu+1)_n}{n!(\mu+\nu+1)_n} \frac{\mu+\nu+1}{2n+\mu+\nu+1} F_n := G_n F_n, \quad (12)$$

Then the recursion relation (4) written for $\{P_n\}$ becomes identical to that of the polynomial $\tilde{H}_n^{(\mu,\nu)}(z;\gamma,\theta)$ shown in Appendix A as Eq. (A8) with the following argument and parameters

$$z^2 = \frac{1}{C(C-\varepsilon)}, \quad \gamma^2 = \frac{1}{16}, \quad \cosh\theta = \frac{\varepsilon - 2C}{\varepsilon}. \quad (13)$$

Consequently, the kth bound state wavefunction with energy $E_k = \varepsilon_k/a^2$ is written as the following finite series

$$\psi_k(x) = f_0(z_k)(x/a)^{\mu+\frac{1}{2}}\left[(x/a)^2 + 2\right]^{\frac{\nu+1}{2}} \sum_{n=0}^{N} G_n^{-1} \tilde{H}_n^{(\mu,\nu)}\left(z_k; \tfrac{1}{4}, \theta_k\right) Q_n^{(\mu,\nu)}(y), \quad (14)$$

where $\mu = \sqrt{2A+\tfrac{1}{4}}$, $\nu = -\sqrt{2B+1}$, $G_n$ is defined in (12), $z_k$ and $\theta_k$ are defined in (13). Therefore, once the energy $E_k$ is obtained, the finite series (14) gives the corresponding bound state. Now, the physical properties of the system, such its energy spectrum, is obtained from the analytic properties of the TRA polynomial $\tilde{H}_n^{(\mu,\nu)}(z;\gamma,\theta)$ such as its weight function, generating function, zeros, asymptotics, etc. Unfortunately, these properties are not yet known and remains an open problem in orthogonal polynomials [10,11]. Therefore, we had to resort to numerical means to obtain the bound states energy spectrum. Table 1 shows the full energy spectrum of $V_1(x)$ for the given set of values of the potential parameters. We used two numerical methods:

(1) The Lagrange mesh method (LMM) parametrized by a linear grid of size $M$ and a variational scale parameter $h$. Please refer to Appendix B.2 and references [12,13] for more details.

(2) Hamiltonian matrix diagonalization (HMD) in a complete Laguerre basis as explained in Appendix B below.

Figure 3a is a plot of the un-normalized bound states wavefunctions corresponding to the energy spectrum in Table 1.

## 3. TRA solution of the potential model (1b)

We repeat the same treatment in section 2 above for the potential model $V_{\text{II}}(x)$. However, we choose the following coordinate transformation and basis parameters

$$y(x) = 2\left[(x/a)+1\right]^2 - 1, \qquad 2\alpha = \mu+1, \qquad 2\beta = -\nu - \tfrac{1}{2}. \tag{15}$$

Subsequently, the action of the wave operator on the basis elements, $\mathcal{D}|\phi_n\rangle$, becomes

$$\begin{aligned}\mathcal{D}|\phi_n(y)\rangle = -\frac{2}{a^2}\frac{(y-1)^{\alpha-1}}{(y+1)^\beta}\Bigg\{&\frac{\mu^2-1}{y-1}-\frac{\nu^2-\tfrac{1}{4}}{y+1}+\frac{1}{2}(2n+\mu+\nu+1)^2-\frac{1}{8}\\&-\frac{2A}{y-1}+\frac{2B}{y+1}-C+\frac{\varepsilon}{2}(y-1)\Bigg\}Q_n^{(\mu,\nu)}(y)\end{aligned} \tag{16}$$

where we have used the differential equation of the Jacobi polynomial (A2). The TRA constraint (3) and the recursion relation of the Jacobi polynomials (A3) dictate that we assign the following values to the Jacobi polynomial parameters

$$\mu^2 = 2A+1, \qquad \nu^2 = 2B + \tfrac{1}{4}. \tag{17}$$

Reality dictates that $A \geq -\tfrac{1}{2}$ and $B \geq -\tfrac{1}{8}$. Moreover, the polynomial parameters inequalities $\mu > -1$ and $\mu+\nu < -2N-1$ dictate that $\mu = \sqrt{2A+1}$ and $\nu = -\sqrt{2B+\tfrac{1}{4}}$. It also shows that the maximum number of bound states that could be obtained by the TRA solution, which is $N+1$, becomes $\left\lfloor \tfrac{1}{2}\left(\sqrt{2B+\tfrac{1}{4}}-\sqrt{2A+1}-1\right)\right\rfloor + 1$. With these choices of basis parameters, Eq. (16) becomes

$$\mathcal{D}|\phi_n\rangle = -\frac{1}{a^2}\frac{(y-1)^{\alpha-1}}{(y+1)^\beta}\left[(2n+\mu+\nu+1)^2 - \frac{1}{4} - (2C+\varepsilon) + \varepsilon y\right]Q_n^{(\mu,\nu)}(y). \tag{18}$$

Using the three-term recursion relation for the Jacobi polynomials (A3) in this equation and comparing the result to the TRA constraint (3), we obtain

$$W(y) = \frac{-E}{y-1}, \tag{19a}$$

$$d_n = \left[(2n+\mu+\nu+1)^2 - \frac{1}{4} - (2C+\varepsilon)\right]\frac{1}{\varepsilon} + \frac{\nu^2-\mu^2}{(2n+\mu+\nu)(2n+\mu+\nu+2)}, \tag{19b}$$

$$b_n = \frac{2(n+1)(n+\mu+\nu+1)}{(2n+\mu+\nu+1)(2n+\mu+\nu+2)}, \qquad c_n = \frac{2(n+\mu+1)(n+\nu+1)}{(2n+\mu+\nu+2)(2n+\mu+\nu+3)}. \tag{19c}$$

Consequently, the recursion relation (4) written in terms of $\{P_n\}$ defined in (12) becomes identical to (A8) of the TRA polynomial $\tilde{H}_n^{(\mu,\nu)}(z;\gamma,\theta)$ with the following parameter and argument relations

$$z^2 = \frac{4}{C(C+\varepsilon)}, \qquad \gamma^2 = \frac{1}{16}, \qquad \cosh\theta = \frac{\varepsilon+2C}{-\varepsilon}. \tag{20a}$$

The condition that $\cosh\theta \geq 1$ dictates that $-C < \varepsilon < 0$, which is the same condition that guarantees reality of $z$. On the other hand, for $\varepsilon < -C$ we compare the recursion relation (4) written for $\{P_n\}$ to Eq. (A7) and we conclude that $P_n = H_n^{(\mu,\nu)}(z;\gamma,\theta)$ with

$$z^2 = \frac{-4}{C(C+\varepsilon)}, \qquad \gamma^2 = \frac{1}{16}, \qquad \cos\theta = \frac{\varepsilon + 2C}{-\varepsilon}. \tag{20b}$$

Finally, the kth bound state wavefunction with energy $E_k = \varepsilon_k/a^2$ is written as the following finite series for $-C < \varepsilon_k < 0$

$$\psi_k(x) = 2^{\frac{\mu+\nu}{2}+\frac{3}{4}} f_0(z_k)\left(\tfrac{x}{a}+1\right)^{\nu+\frac{1}{2}}\left[\left(\tfrac{x}{a}\right)^2 + 2\tfrac{x}{a}\right]^{\frac{\mu+1}{2}} \sum_{n=0}^{N} G_n^{-1} \tilde{H}_n^{(\mu,\nu)}\left(z_k;\tfrac{1}{4},\theta_k\right) Q_n^{(\mu,\nu)}(y). \tag{21a}$$

On the other hand, for $\varepsilon_k < -C$, the wavefunction becomes

$$\psi_k(x) = 2^{\frac{\mu+\nu}{2}+\frac{3}{4}} f_0(z_k)\left(\tfrac{x}{a}+1\right)^{\nu+\frac{1}{2}}\left[\left(\tfrac{x}{a}\right)^2 + 2\tfrac{x}{a}\right]^{\frac{\mu+1}{2}} \sum_{n=0}^{N} G_n^{-1} H_n^{(\mu,\nu)}\left(z_k;\tfrac{1}{4},\theta_k\right) Q_n^{(\mu,\nu)}(y), \tag{21b}$$

where $\mu = \sqrt{2A+1}$, $\nu = -\sqrt{2B+\tfrac{1}{4}}$, $G_n$ is defined in (12), $z_k$ and $\theta_k$ are defined in (20a) and (20b), respectively. Now, the physical properties of the system is obtained from those of the TRA polynomials $H_n^{(\mu,\nu)}(z;\gamma,\theta)$ and $\tilde{H}_n^{(\mu,\nu)}(z;\gamma,\theta)$, which is still an open problem in orthogonal polynomials. Table 2 shows the full energy spectrum of $V_{II}(x)$ for the given set of values of the potential parameters obtained using LMM and HMD. Comparison of the results listed in Table 1 and Table 2 shows that our calculation with $V_{II}(x)$ is far less accurate than with $V_I(x)$. We believe that, most likely, this calculation difficulty is due to the long range $1/x$ singularity present in $V_{II}(x)$ but not in $V_I(x)$. To give a pictorial representation demonstrating this difficulty, we show in Figure 4 the two potential plots with the energy spectrum super-imposed. One can observe two features of the difficulty: (1) how high the ground state from the bottom of the potential $V_{II}(x)$ compared to $V_I(x)$, and (2) how rapid the reduction in the energy spacing of the $V_{II}(x)$ spectrum compared to $V_I(x)$. Figure 3b is a plot of the un-normalized bound states wave functions corresponding to the energy spectrum shown in Table 2.

## 4. Conclusion

The potential plots in Figure 1 and the spectral phase diagram in Figure 2 show that the two potential functions introduced in this work have a rich structure. Therefore, they could be used to model various physical systems with wide range of structural and dynamical properties. These two potentials do not belong to the class of exactly solvable quantum mechanical problems. Nonetheless, we were able to use the tridiagonal representation approach and obtain exact bound states solutions with the constraint that all potential parameters are positive. That is, the TRA solution space is confined to the green "B&R" region of the SPD. We should also mention that for $C<0$, the TRA can also be used to obtain only the lowest $\left\lfloor -\frac{\mu+\nu+1}{2} \right\rfloor + 1$ bound states. In the SPD, these solutions lie in the blue "B" region.

The shortcoming of our solution is that the analytic properties of the TRA polynomials $H_n^{(\mu,\nu)}(z;\gamma,\theta)$ and $\tilde{H}_n^{(\mu,\nu)}(z;\gamma,\theta)$ that contain all the physical properties of the system are yet to be derived. This is a mathematical problem, which goes beyond the scope of this work and expertise of the authors. However, for the complete descriptions of the solution given by Eq. (14) and Eq. (21), one needs only the of the corresponding energy eigenvalue $E_k$. We used two independent numerical routines to obtain highly accurate evaluation of the complete energy spectrum as shown in the Tables.

## Appendix A: Jacobi polynomial on the semi-infinite line

For ease of reference, we reproduce Appendix A in our recent work [14] where these polynomials are defined over the semi-infinite interval $y \geq 1$. The conventional Jacobi polynomials $P_n^{(\mu,\nu)}(y)$ are defined on the finite interval $-1 \leq y \leq +1$ and to distinguish them from $P_n^{(\mu,\nu)}(y)$ we use the notation $Q_n^{(\mu,\nu)}(y)$:

$$Q_n^{(\mu,\nu)}(y) = \frac{\Gamma(n+\mu+1)}{\Gamma(n+1)\Gamma(\mu+1)} {}_2F_1\left(\begin{matrix}-n, n+\mu+\nu+1\\ \mu+1\end{matrix}\bigg|\frac{1-y}{2}\right) = (-1)^n Q_n^{(\nu,\mu)}(-y). \tag{A1}$$

where $n = 0, 1, 2, ..., N$, $\mu > -1$ and $\mu + \nu < -2N - 1$. It satisfies the following differential equation

$$\left\{(y^2-1)\frac{d^2}{dy^2} + \left[(\mu+\nu+2)y+\mu-\nu\right]\frac{d}{dy} - n(n+\mu+\nu+1)\right\}Q_n^{(\mu,\nu)}(y) = 0, \tag{A2}$$

It also satisfies the following three-term recursion relation

$$yQ_n^{(\mu,\nu)}(y) = \frac{\nu^2-\mu^2}{(2n+\mu+\nu)(2n+\mu+\nu+2)}Q_n^{(\mu,\nu)}(y)$$
$$+ \frac{2(n+\mu)(n+\nu)}{(2n+\mu+\nu)(2n+\mu+\nu+1)}Q_{n-1}^{(\mu,\nu)}(y) + \frac{2(n+1)(n+\mu+\nu+1)}{(2n+\mu+\nu+1)(2n+\mu+\nu+2)}Q_{n+1}^{(\mu,\nu)}(y) \tag{A3}$$

and the following differential relation

$$(y^2-1)\frac{d}{dy}Q_n^{(\mu,\nu)} = 2(n+\mu+\nu+1)\left[\frac{(\nu-\mu)n}{(2n+\mu+\nu)(2n+\mu+\nu+2)}Q_n^{(\mu,\nu)}\right.$$
$$\left. - \frac{(n+\mu)(n+\nu)}{(2n+\mu+\nu)(2n+\mu+\nu+1)}Q_{n-1}^{(\mu,\nu)} + \frac{n(n+1)}{(2n+\mu+\nu+1)(2n+\mu+\nu+2)}Q_{n+1}^{(\mu,\nu)}\right] \tag{A4}$$

The associated orthogonality relation reads as follows

$$\int_1^\infty (y-1)^\mu (y+1)^\nu Q_n^{(\mu,\nu)}(y)Q_m^{(\mu,\nu)}(y)dy = \frac{2^{\mu+\nu+1}}{2n+\mu+\nu+1}\frac{\Gamma(n+\mu+1)\Gamma(n+\nu+1)}{\Gamma(n+1)\Gamma(n+\mu+\nu+1)}\frac{\sin\pi\nu}{\sin\pi(\mu+\nu+1)}\delta_{nm}, \tag{A5}$$

where $n, m \in \{0, 1, 2, ..., N\}$. Equivalently (see Eq. 4.9 in Ref. [15]),

$$\int_1^\infty (y-1)^\mu (y+1)^\nu Q_n^{(\mu,\nu)}(y)Q_m^{(\mu,\nu)}(y)dy = \frac{(-1)^{n+1}2^{\mu+\nu+1}}{2n+\mu+\nu+1}\frac{\Gamma(n+\mu+1)\Gamma(n+\nu+1)\Gamma(-n-\mu-\nu)}{\Gamma(n+1)\Gamma(-\nu)\Gamma(\nu+1)}\delta_{nm}. \tag{A6}$$

The TRA polynomial $H_n^{(\mu,\nu)}(z;\alpha,\theta)$ is defined in Ref. [10] by its three-term recursion relation, Eq. (8) therein, which we rewrite here as

$$(\cos\theta)H_n^{(\mu,\nu)}(z;\gamma,\theta) = \left\{\left[\left(n+\frac{\mu+\nu+1}{2}\right)^2-\gamma^2\right]z(\sin\theta)-\frac{\nu^2-\mu^2}{(2n+\mu+\nu)(2n+\mu+\nu+2)}\right\}H_n^{(\mu,\nu)}(z;\gamma,\theta) \quad \text{(A7)}$$
$$-\frac{2(n+\mu)(n+\nu)}{(2n+\mu+\nu)(2n+\mu+\nu+1)}H_{n-1}^{(\mu,\nu)}(z;\gamma,\theta)-\frac{2(n+1)(n+\mu+\nu+1)}{(2n+\mu+\nu+1)(2n+\mu+\nu+2)}H_{n+1}^{(\mu,\nu)}(z;\gamma,\theta),$$

where $H_0^{(\mu,\nu)}(z;\alpha,\theta)=1$ and $H_{-1}^{(\mu,\nu)}(z;\alpha,\theta):=0$. For some range of values of the polynomial parameters, it is more appropriate to define $\tilde{H}_n^{(\mu,\nu)}(z;\gamma,\theta)=H_n^{(\mu,\nu)}(-iz;\alpha,i\theta)$, which maps the recursion (A7) into

$$(\cosh\theta)\tilde{H}_n^{(\mu,\nu)}(z;\gamma,\theta) = \left\{\left[\left(n+\frac{\mu+\nu+1}{2}\right)^2-\gamma^2\right]z(\sinh\theta)-\frac{\nu^2-\mu^2}{(2n+\mu+\nu)(2n+\mu+\nu+2)}\right\}\tilde{H}_n^{(\mu,\nu)}(z;\gamma,\theta) \quad \text{(A8)}$$
$$-\frac{2(n+\mu)(n+\nu)}{(2n+\mu+\nu)(2n+\mu+\nu+1)}\tilde{H}_{n-1}^{(\mu,\nu)}(z;\gamma,\theta)-\frac{2(n+1)(n+\mu+\nu+1)}{(2n+\mu+\nu+1)(2n+\mu+\nu+2)}\tilde{H}_{n+1}^{(\mu,\nu)}(z;\gamma,\theta),$$

## Appendix B: Energy spectrum calculations

### B.1. Hamiltonian matrix diagonalization in the Laguerre basis

For this calculation, we choose the following complete basis

$$\chi_m(x) = C_m (z)^{\frac{\sigma+1}{2}} e^{-z/2} L_m^\sigma(z), \quad \text{(B1)}$$

where $z=\lambda x$, $L_m^\sigma(z)$ is the Laguerre polynomial, and $C_m = \sqrt{m!/\Gamma(m+\sigma+1)}$. The positive scale parameter $\lambda$ is an optimization parameter with inverse length dimension. The action of the kinetic energy operator on the basis (B1) is as follow

$$-\frac{1}{2}\frac{d^2\chi_m}{dx^2} = -\frac{\lambda^2}{2}C_m(z)^{\frac{\sigma-1}{2}}e^{-z/2}\left[-\frac{1}{2}(2n+\sigma+1)+\frac{\sigma^2-1}{4z}+\frac{z}{4}\right]L_m^\sigma(z), \quad \text{(B2)}$$

where we have used the differential equation of the Laguerre polynomial. To obtain the action of the Hamiltonian operator on the basis (B1), we add to (B2) the following potential energy component $V_1(x)\chi_m(x)$

$$V_1\chi_m = \lambda^2 C_m(z)^{\frac{\sigma+1}{2}}e^{-z/2}\left\{\frac{A}{z^2}+\frac{2C-A}{z^2+2(\lambda a)^2}-\frac{2(\lambda a)^2 B}{[z^2+2(\lambda a)^2]^2}\right\}L_m^\sigma(z). \quad \text{(B3)}$$

To eliminate the singular $z^{-1}$ term inside the square brackets of (B2), we match it with the singular $z^{-2}$ term inside the curly brackets of (B2) by choosing $\sigma^2=1+8A$. Using the recursion relation and orthogonality of the Laguerre polynomials, the Hamiltonian matrix in the basis (B1) with $\sigma=\sqrt{1+8A}$ becomes

$$\langle \chi_n | H | \chi_m \rangle = \frac{\lambda^2}{8} \left[ (2n+\sigma+1)\delta_{n,m} + \delta_{n,m+1}\sqrt{n(n+\sigma)} + \delta_{n,m-1}\sqrt{(n+1)(n+\sigma+1)} \right]$$
$$+ \lambda^2 (2C-A) \langle n | \frac{z}{z^2+2(\lambda a)^2} | m \rangle - 2a^2\lambda^4 B \langle n | \frac{z}{[z^2+2(\lambda a)^2]^2} | m \rangle \tag{B4}$$

where we have defined

$$\langle n | g(z) | m \rangle = C_n C_m \int_0^\infty (z)^\sigma e^{-z} g(z) L_n^\sigma(z) L_m^\sigma(z) dz, \tag{B5}$$

We can use Gauss quadrature associated with the Laguerre polynomial to get a very good approximation for this integral. Moreover, the basis overlap matrix is $\Omega_{n,m} = \langle \chi_n | \chi_m \rangle = (2n+\sigma+1)\delta_{n,m} - \delta_{n,m+1}\sqrt{n(n+\sigma)} - \delta_{n,m-1}\sqrt{(n+1)(n+\sigma+1)}$. The accuracy in the evaluation of the two integrals in (B4) improves with the size of the basis $\{\chi_m(x)\}_{m=0}^M$ and with an optimized choice for the scale parameter $\lambda$. Finally, with the matrices $H$ and $\Omega$ being determined, we can obtain the energy spectrum $\{E_k\}$ from the generalized eigenvalue equation $H|\psi_k\rangle = E_k \Omega |\psi_k\rangle$. In Table 1, we list these results for the given basis size $M$ and optimization parameter $\lambda$.

For the potential $V_{II}(x)$, we repeat the same procedure in the same basis (B1) but with $\sigma = \sqrt{1+2A}$. Consequently, we obtain the following matrix elements of the Hamiltonian

$$\langle \chi_n | H | \chi_m \rangle = \frac{\lambda^2}{8} \left[ (2n+\sigma+1)\delta_{n,m} + \delta_{n,m+1}\sqrt{n(n+\sigma)} + \delta_{n,m-1}\sqrt{(n+1)(n+\sigma+1)} \right]$$
$$- \frac{\lambda^2}{2}(A+2B-2C) \langle n | \frac{1}{z+2\lambda a} | m \rangle + \lambda^2 \frac{A}{4} \langle n | \frac{z}{(z+2\lambda a)^2} | m \rangle + \lambda^2 B \langle n | \frac{z}{(z+\lambda a)^2} | m \rangle \tag{B6}$$

The energy spectrum $\{E_k\}$ is obtained from the generalized eigenvalue equation $H|\psi_k\rangle = E_k \Omega |\psi_k\rangle$. In Table 2, we list these results for the given basis size $M$ and optimization parameter $\lambda$.

### B.2. Lagrange mesh method

For this calculation, we use the Lagrange mesh method (LMM) based on Gauss quadrature associated with the Laguerre polynomial [12,13]. Starting with Schrödinger equation (6), and using the Lagrange-Laguerre basis [12],

$$\varphi_i(x) = \frac{(-1)^i x}{\sqrt{x_i}(x-x_i)} e^{-x/2} L_M(x), \tag{B7}$$

with $i = 1, 2, .., M$ and $x_i$ being one of the (dimensionless) zeros of the Laguerre polynomial $L_M(x)$. Consequently, the wave equation reduces to the following generalized eigenvalue equation

$$\left[ \frac{1}{2h^2} \vec{T} + \vec{V}_h \right] |\zeta\rangle = E(\Xi|\zeta\rangle), \tag{B8}$$

where

$$\left(\vec{\vec{T}}\right)_{ij} = \begin{cases} \dfrac{1}{12x_i}\left[4+(4M+2)x_i - x_i^2\right] - \dfrac{1}{4}S_{ii} & , i = j \\[2ex] \left[\dfrac{x_i + x_j}{(x_i - x_j)^2} - \dfrac{1}{4}\right]S_{ij} & , i \neq j \end{cases} \quad \text{(B9)}$$

with $S_{ij} = (-1)^{i-j}/\sqrt{x_i x_j}$, $\left(\vec{\vec{V}}_h\right)_{ij} = V(hx_i)\delta_{ij}$, and $\Xi_{ij} = \delta_{ij} + S_{ij}$ is the basis overlap matrix [12]. The variational parameter $h$ is chosen within a plateau where the eigenvalues are stable. In our case we took $h = 0.001$ and $M = 3000$.

## Tables Captions

**Table 1.** The negative of the complete energy spectrum (in atomic units) for the potential $V_I(x)$ obtained using LMM and HMD techniques for the potential parameter values: $\{a, A, B, C\} = \{1,1,100,2\}$. For the LMM, we took the variational parameter $h = 0.1$ and a grid size $M = 50$ whereas for the HMD, we took the scale parameter $\lambda = 10$ and a matrix size $M = 100$.

**Table 2.** The negative of the complete energy spectrum (in atomic units) for the potential $V_{II}(x)$ obtained using LMM and HMD techniques for the potential parameter values: $\{a, A, B, C\} = \{1,1,100,2\}$. For the LMM, we took the variational parameter $h = 0.001$ and a grid size $M = 3000$ whereas for the HMD, we took the scale parameter $\lambda = 15$ and a matrix size $M = 100$.

## Figures Captions

**Fig. 1.** The two potential functions $V_I(x)$ and $V_{II}(x)$ (in units of *A*) for several values of the associated parameters: (a) $V_I(x)$ with $C = 5A$ and several values of *B*, (b) $V_{II}(x)$ with $C = 2A$ and several values of *B*.

**Fig. 2.** The spectral phase diagram (SPD) for: (a) $V_I(x)$, and (b) $V_{II}(x)$. The diagram shows the distribution of the energy spectrum (scattering states "S", bound states "B", and resonances "R") as a function of the potential parameters.

**Fig. 3.** The un-normalized wavefunctions corresponding to the energy spectrum of: (a) $V_I(x)$ in Table 1, and (b) $V_{II}(x)$ in Table 2. The horizontal *x*-axis is in units of *a*.

**Fig. 4.** The two potential functions with their energy spectra superimposed. We used the physical parameters and results of Table 1 and 2. One can observe: (1) how high the ground state from the bottom of the potential $V_{II}(x)$ compared to $V_I(x)$, and (2) how rapid the reduction in the energy spacing of the $V_{II}(x)$ spectrum compared to $V_I(x)$.

**Table 1**

| $n$ | HMD | LMM |
|---|---|---|
| 0 | 26.92691153111182 | 26.92691153111209 |
| 1 | 14.73315268981235 | 14.73315268981183 |
| 2 | 6.57626497755787 | 6.57626497755784 |
| 3 | 1.98828286255332 | 1.98828286255328 |
| 4 | 0.18985650550822 | 0.18985650519293 |

**Table 2**

| $n$ | HMD | LMM |
|---|---|---|
| 0 | 535.330051916 | 535.330051482 |
| 1 | 120.017539122 | 120.017539017 |
| 2 | 30.767199571 | 30.767199548 |
| 3 | 6.397103500 | 6.397103493 |
| 4 | 0.610381100 | 0.610381099 |

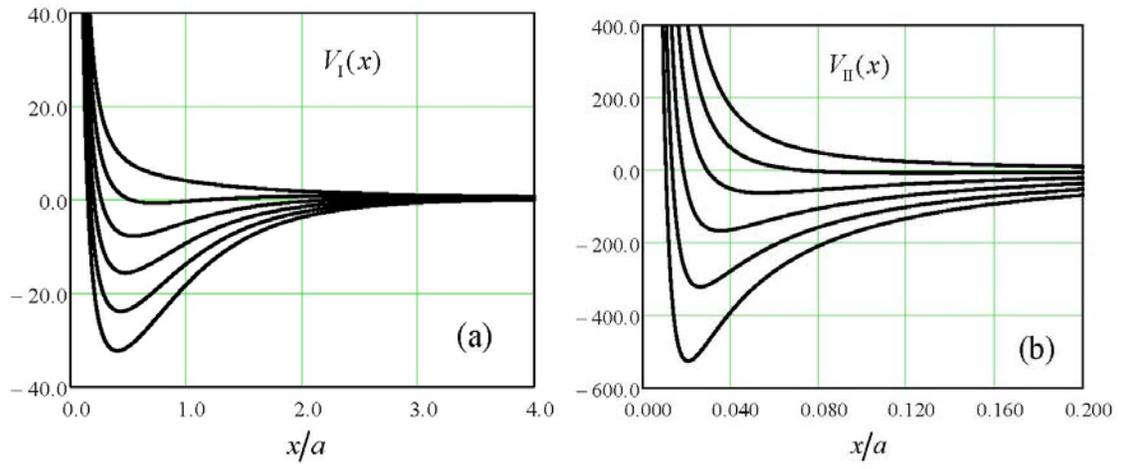

**Fig. 1**

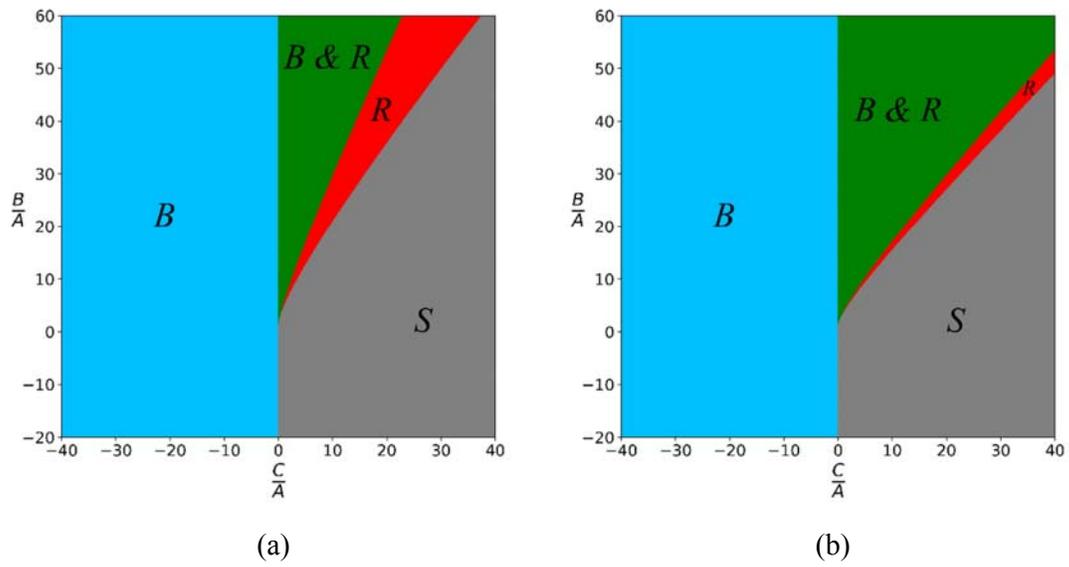

(a)                                                                 (b)

**Fig. 2**

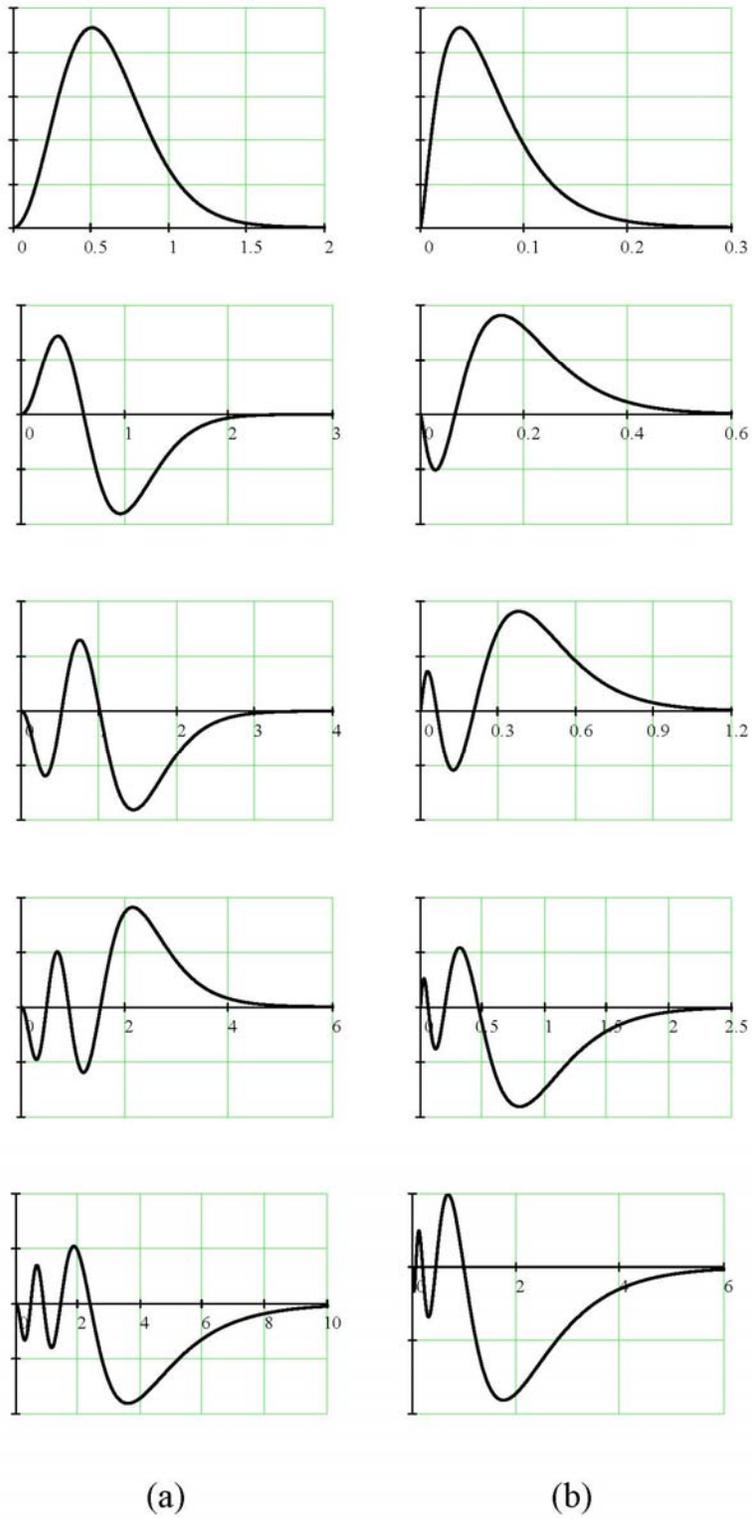

**Fig. 3**

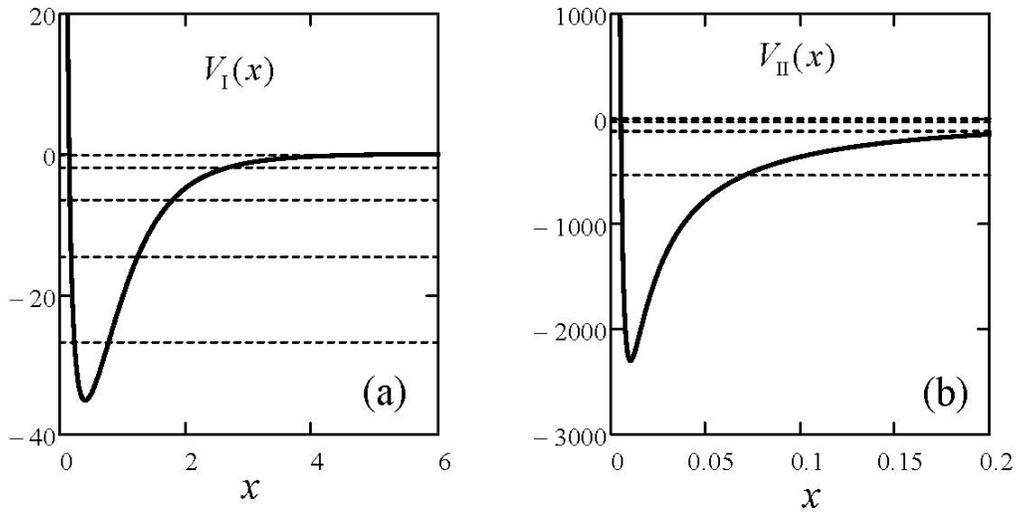

**Fig. 4**